\documentclass[preprint, superscriptaddress, prb, floatfix]{revtex4-2}
\usepackage{lineno}
\usepackage{graphicx}
\usepackage{float}
\usepackage{amssymb}
\usepackage{mathrsfs}
\usepackage{physics}
\usepackage{hyperref}
\usepackage{booktabs}
\usepackage[british]{babel}
\usepackage{caption}
\usepackage{bm}
\newcommand{\opa}{\hat{a}}
\newcommand{\opad}{\hat{a}^\dagger}

\newcommand{\oph}{\hat{H}}

\newcommand{\opu}{\hat{U}}
\newcommand{\opud}{\hat{U}^\dagger}

\begin{document}


\title{Acoustic phonon phase gates with number-resolving phonon detection}
\author{Hong Qiao}
\affiliation{Pritzker School of Molecular Engineering, University of Chicago, Chicago IL 60637, USA}

\author{Zhaoyou Wang}
\affiliation{Pritzker School of Molecular Engineering, University of Chicago, Chicago IL 60637, USA}

\author{Gustav Andersson}
\affiliation{Pritzker School of Molecular Engineering, University of Chicago, Chicago IL 60637, USA}

\author{Alexander Anferov}
\affiliation{Pritzker School of Molecular Engineering, University of Chicago, Chicago IL 60637, USA}

\author{Christopher R. Conner}
\affiliation{Pritzker School of Molecular Engineering, University of Chicago, Chicago IL 60637, USA}

\author{Yash J. Joshi}
\affiliation{Pritzker School of Molecular Engineering, University of Chicago, Chicago IL 60637, USA}

\author{Shiheng Li}
\affiliation{Pritzker School of Molecular Engineering, University of Chicago, Chicago IL 60637, USA}
\affiliation{Department of Physics, University of Chicago, Chicago IL 60637, USA}

\author{Jacob M. Miller}
\affiliation{Pritzker School of Molecular Engineering, University of Chicago, Chicago IL 60637, USA}
\affiliation{Department of Physics, University of Chicago, Chicago IL 60637, USA}

\author{Xuntao Wu}
\affiliation{Pritzker School of Molecular Engineering, University of Chicago, Chicago IL 60637, USA}

\author{Haoxiong Yan}
\affiliation{Pritzker School of Molecular Engineering, University of Chicago, Chicago IL 60637, USA}

\author{Liang Jiang}
\affiliation{Pritzker School of Molecular Engineering, University of Chicago, Chicago IL 60637, USA}

\author{Andrew N. Cleland}
\email{anc@uchicago.edu}
\affiliation{Pritzker School of Molecular Engineering, University of Chicago, Chicago IL 60637, USA}
\affiliation{Center for Molecular Engineering and Material Science Division, Argonne National Laboratory, Lemont IL 60439, USA}

\maketitle

\textbf{Linear optical quantum computing (LOQC) provides a compelling approach to quantum information processing, with a short list of physical requirements; however, experimental implementations have faced significant challenges \cite{Knill2001, Kok2007, Maring2024}. Itinerant phonons in quantum acoustics, combined with superconducting qubits, offer a compelling alternative to the quantum optics approach \cite{Qiao2023}. Here we demonstrate key advances in the ability to manipulate and measure acoustic phonon quantum states: First, we demonstrate deterministic phase control of itinerant one- and two-phonon qubit states, measured using an acoustic Mach-Zehnder interferometer. We implement phonon phase control using the frequency-dependent scattering of phonon states from a superconducting transmon qubit. The acoustic interferometer used to measure the resulting phonon phase achieves a noise-floor-limited Hong-Ou-Mandel (HOM) interference visibility of 98.1\%, representing a significant improvement over our previous demonstration \cite{Qiao2023}. Additionally, we propose and implement a multi-phonon detection scheme that enables coherent conversion between itinerant one- and two-phonon Fock states and transmon qutrit states, transforming for example the Hong-Ou-Mandel two-phonon entangled output state $|02\rangle - |20\rangle$ into the entangled state of two transmons. The tight integration of quantum acoustics with superconducting circuits native to our implementation promises further advances, including deterministic phonon quantum gates \cite{Wang2024} with direct applications to quantum computing \cite{Zhong2020, Madsen2022, AghaeeRad2025, Alexander2025}.}

Optical quantum computing architectures, which use photons as quantum information carriers, are under active development, including through the large-scale integration of photonic elements on a manufacturable platform \cite{Knill2001, Kok2007, Bogaerts2020, Zhong2020, Madsen2022,  AghaeeRad2025, Alexander2025}. Microwave-frequency acoustic phonons offer an interesting alternative to photons \cite{OConnell2010, Chan2011, Chu2017, Satzinger2018, Qiao2023}, featuring compact dimensions afforded by the low speed of sound \cite{ArrangoizArriola2019, Fu2019, Kuzyk2018, Taylor2022}, as well as potentially long lifetimes \cite{Chu2018, MacCabe2020, Zivari2022, Bozkurt2023, Bozkurt2024}.  There are also compelling applications to microwave-to-optical transduction \cite{Bochmann2013, Andrews2014, Mirhosseini2020, Jiang2020, Arnold2020, Weaver2023} as well as quantum information storage \cite{Hann2019, Wallucks2020, Chamberland2022, Wollack2022, Chou2025}. Among the various quantum acoustics platforms, itinerant surface acoustic wave (SAW) phonons \cite{Gustafsson2014, Bienfait2019, Dumur2021} provide unique opportunities, due to their well-defined propagating spatial and temporal modes, similar to optical photons, and the deterministic and quantum-coherent single-phonon emission and detection afforded by their integration with superconducting qubits \cite{Bienfait2019}.  

Optical phase shifters are essential building blocks for optical quantum computing, where programmable phase control of photons can be implemented using e.g. thermo-optical phase shifters \cite{VanCampenhout2010,Bogaerts2020}, photonic light-matter interfaces \cite{Englund2007, Tiecke2014, Bhaskar2020}, or microwave photon-circuit QED systems \cite{Kono2018, Besse2018, Zhang2020}. The analogous precise and efficient phase control of acoustic phonons would provide similar functionality. In previous work, single-phonon phase control was indirectly realized by capturing a phonon in a transmon qubit and using single qubit gates prior to re-emitting the phonon \cite{Qiao2023}. Here we demonstrate a new method for direct phase control of both one- and two-phonon states, achieved by scattering the phonon state from a transmon qubit in a simple and efficient process, validated using a balanced acoustic Mach-Zehnder interferometer. We further propose and implement a method for the coherent capture of itinerant one- and two-phonon states by a transmon qutrit, providing an important extension of our prior single-phonon capture capability and promising further extensions for two-phonon state control.

\begin{figure}[t]
\begin{center}
	\includegraphics[width=1\textwidth]{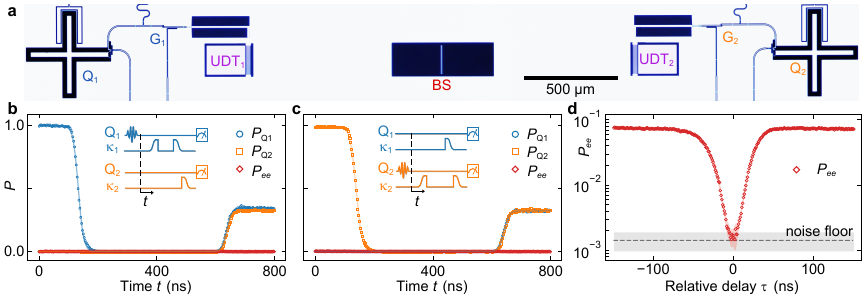}
	\caption{
    \label{fig1}
    {\bf Device, characterization and phonon indistinguishability.}
    \textbf{a,} Optical micrograph of device, comprising two superconducting qubits ($\text{Q}_1$ and $\text{Q}_2$) coupled to unidirectional transducers ($\text{UDT}_1$ and $\text{UDT}_2$) via tunable couplers ($\text{G}_1$ and $\text{G}_2$), on either side of an acoustic channel with a mid-point phonon beamsplitter (BS). Qubits and couplers are fabricated on a sapphire die, with UDTs and BS on a separate, smaller lithium niobate die; the micrograph is imaged through the backside of the flip-chip assembled device. \textbf{b, c,} A single phonon is sent from $\text{Q}_1$ ($\text{Q}_2$) towards the BS, where it is ``split'' and the BS output subsequently captured by each qubit. Simultaneous measurements of both qubits are made at time $t$, yielding the excited state populations $P_\text{Q1}$ (blue circles),  $P_\text{Q2}$ (orange squares), and the joint qubit excitation probability $P_{ee}$ (red diamonds). Solid color lines are numerical simulations. Insets: Control pulse sequences. \textbf{d,} Two-qubit joint excitation probability $P_{ee}$ measured as a function of relative phonon delay $\tau$, showing a pronounced Hong-Ou-Mandel dip for coincident phonons, with a visibility $\mathcal{V}_\text{HOM} = 0.981 \pm 0.007$. Light red shaded area denotes uncertainty, calculated from ten repeated scans for each delay $\tau$. Dashed gray line and gray shaded area indicate the noise floor for $P_{ee}$ and its uncertainty, respectively. All uncertainties are one standard deviation.}
\end{center}
\end{figure}

Our device, shown in Fig.~\ref{fig1}{\bf a}, is carefully optimized over previous designs \cite{Qiao2023}, primarily to center the phonon beamsplitter in the 2 mm-long acoustic channel as well as achieve a balanced 1:1 transmission:reflection ratio.  The device comprises two superconducting Xmon qubits $\text{Q}_1$ and $\text{Q}_2$ \cite{Koch2007, Barends2013}, coupled via two tunable couplers $\text{G}_1$ and $\text{G}_2$ \cite{Chen2014} to two unidirectional interdigitated transducers $\text{UDT}_1$ and $\text{UDT}_2$ \cite{Dumur2021}; the design enables 10 ns phonon emission times at maximum coupling (see Extended Data Fig.~\ref{figS3}). The device is operated in a dilution refrigerator with a base temperature of 7 mK. 

We characterize the device by performing single-phonon beamsplitter experiments: We excite one qubit (Q$_1$ or Q$_2$) to its $|e\rangle$ state, then emit a phonon at a center frequency of 3.925 GHz with a hyperbolic secant waveform, $\phi_{1,2}(t) \propto \sech(t/\sigma_{1,2})$, with a wavepacket full width at half maximum (FWHM) of $2.6\times\sigma_{1,2} \approx 52$ ns, the waveform set by calibrated control of the qubit-coupler-UDT coupling strength. The emitted phonon is split by the beamsplitter, then captured by each qubit using a timed coupling strength matching that used for phonon release. In Fig.~\ref{fig1}{\bf b} and {\bf c}, we show the excited state probability $P_e(t)$ for each qubit with time $t$, as well as the probability $P_{ee}(t)$ for both qubits to be excited. The nearly-equal transmitted and reflected qubit populations are consistent with a balanced beamsplitter; the zero-valued $P_{ee}$ is also as expected for a single phonon experiment. 

We also perform a Hong-Ou-Mandel (HOM) effect experiment, testing the indistinguishability of the phonons emitted by each qubit. Following release of nominally identical phonons from each qubit, with a programmed time delay $\tau$, the qubits are used to capture the beamsplitter output; due to the qubit anharmonicity, each qubit can catch at most one phonon. We show the resulting two-qubit excitation probability $P_{ee}(\tau)$ in Fig.~\ref{fig1}{\bf d} \cite{Qiao2023}, displaying the expected behavior with delay $\tau$, with two-qubit excitations suppressed near $\tau = 0$ due to the prevalence of the two-phonon entangled output state $(\ket{20}-\ket{02})/\sqrt{2}$, resulting in only one qubit being excited during the catch process. We calculate the visibility of HOM dip as $\mathcal{V}_\text{HOM}\equiv (P_{ee,\text{max}}-P_{ee,\text{min}})/P_{ee,\text{max}}=0.981\pm0.007$, limited by the dual-qubit excitation background. We estimate this noise floor background to be $P_{ee\text{,floor}}=0.0014 \pm 0.0005$ by evaluating the final $P_{ee}$ values in the single-phonon experiments in Fig.~\ref{fig1}{\bf b} (see Extended Data Fig.~\ref{figS1}), which limits the maximum measurable HOM visibility to $(P_{ee,\text{max}}-P_{ee\text{,floor}})/P_{ee,\text{max}}=0.981\pm0.006$.

\begin{figure}[ht]
\begin{center}
\includegraphics[width=0.7\textwidth]{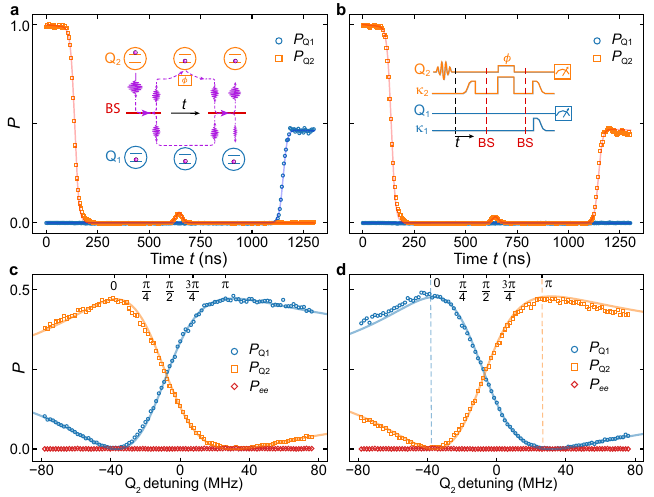}
	\caption{
    \label{fig2}
    {\bf Single phonon phase gate measured by acoustic Mach-Zehnder interferometry.}
    \textbf{a, b, } A single phonon emitted by $\text{Q}_2$ passes through the beamsplitter BS, the BS output then reflecting from both qubit-coupled UDTs. The phonon reflecting from $\text{Q}_2$'s UDT interacts with Q$_2$ via $\text{Q}_2$'s coupler, which is left on with constant coupling during this process. This results in a phase shift $\phi$ determined by $\text{Q}_2$'s $eg$ transition frequency relative to the phonon center frequency. During this process, $\text{Q}_1$'s coupler is left off. The reflections from the two UDTs then interfere at the beamsplitter, whose output is then captured by both qubits. In panel a this results in the phonon being captured by $\text{Q}_1$, in panel b by $\text{Q}_2$. The small and short-lived increase in $P_\text{Q2}$ at around 635 ns is due to the $\text{Q}_2$ scattering process.  \textbf{c, d, } We vary $\text{Q}_2$'s frequency during the scattering process and plot the final two-qubit excited state populations $P_\text{Q1}$, $P_\text{Q2}$ and $P_{ee}$ as a function of the detuning, in panel \textbf{c} (panel \textbf{d}) releasing the initial phonon from $\text{Q}_1$ ($\text{Q}_2$). Top axis shows relative phase difference, referenced to $P_\text{Q1,min}$ ($P_\text{Q2,min}$). Blue and orange vertical dashed lines in panel \textbf{d} correspond to Q$_2$ detunings used in panels \textbf{a} and \textbf{b}, respectively, and correspond to a phase difference of $\pi$. Insets in \textbf{a} and \textbf{b} show the experiment schematic and control pulse sequence, respectively.  Solid blue, orange and red lines are simulations.}
\end{center}
\end{figure}

We demonstrate a deterministic single-phonon phase gate using a frequency-dependent scattering process between a qubit and a phonon (see Methods), using the Mach-Zehnder interferometer to measure the change in phonon phase. The process is outlined in Fig.~\ref{fig2}, with panels \textbf{a} and \textbf{b} showing the time-domain process for generating a phase shift such that, subsequent to interference at the beamsplitter, the phonon is either directed to $\text{Q}_1$ or to $\text{Q}_2$, respectively. In \textbf{a} and \textbf{b} the phonon originates from Q$_2$; similar measurements for Q$_1$-released phonons are shown in Extended Data Fig.~\ref{figS5}. 

Panels \textbf{c} and \textbf{d} show more generally the probability for the phonon to be captured by $\text{Q}_1$ or $\text{Q}_2$ as a function of $\text{Q}_2$'s detuning with respect to the phonon frequency, where changing the detuning results in a different scattering phase. In panel \textbf{c} (\textbf{d}) the phonon originates from $\text{Q}_1$ ($\text{Q}_2$), respectively. This phase-controlled phonon routing has an average visibility of $\mathcal{V}_\text{MZ} \equiv (P_\text{max}-P_\text{min})/(P_\text{max}+P_\text{min}) =  0.979 \pm 0.006$. The single qubit excitation noise floor limits the maximum measurable visibility to $0.986 \pm 0.001$ (see Extended Data Fig.~\ref{figS2}).

\begin{figure}[ht!]
\begin{center}
\includegraphics[width=0.8\textwidth]{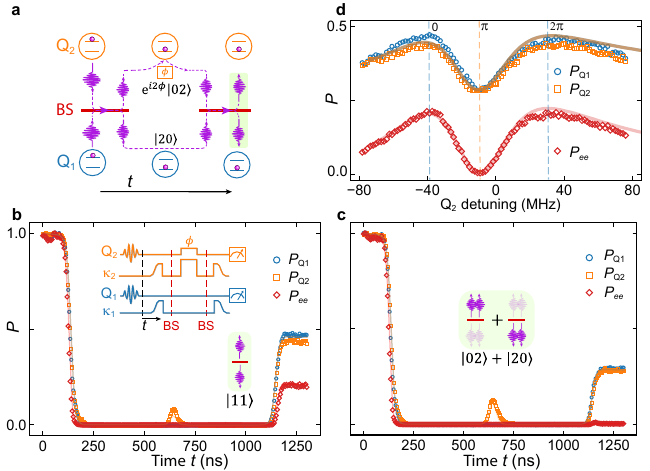}
\captionsetup{width=1\linewidth}
\caption{
\label{fig3}
{\bf Two-phonon phase control.}  
\textbf{a} Schematic of the two-phonon phase control process, similar to Fig.~\ref{fig2}, but here each qubit is excited and releases one phonon to the BS, ideally yielding the HOM output state. \textbf{b, c,} Time traces for $P_\text{Q1}$, $P_\text{Q2}$ and $P_{ee}$, with two-phonon phase selected to yield maximum $P_{ee}$ and minimum $P_{ee}$, respectively, corresponding to beamsplitter output states $\ket{11}$ and $(\ket{02}+\ket{20})/\sqrt{2}$. \textbf{d, } Excited state populations as a function of qubit Q$_2$'s detuning during scattering with respect to the phonon frequency. The maxima in $P_{ee}$ correspond to a $2\phi=2\pi$ phase difference. Left hand blue (maximum $P_{ee}$) and center orange (minimum $P_{ee}$) dashed lines correspond to a $2\phi=\pi$ phase difference, and indicate $\text{Q}_2$'s detuning used in panels \textbf{b} and \textbf{c}, respectively. The asymmetry in $P_{ee}$ with respect to $Q_2$'s detuning is due to a structural phase difference between the two paths of the MZ interferometer. Control pulse sequence is shown inset in \textbf{b}. Solid lines are numerical simulations.}
\end{center}
\end{figure}

We extend the single-phonon phase gate to phase control of a two-phonon state,  shown in Fig.~\ref{fig3}. As shown in Fig.~\ref{fig3}{\bf a}, we create a two-phonon state via the acoustic Hong-Ou-Mandel effect \cite{Qiao2023}, interfering identical phonons from each qubit at the beamsplitter, ideally generating the state $\ket{\psi}_\text{HOM} = (\ket{02}-\ket{20})/\sqrt{2}$. For conditions that give a one-phonon phase shift $\phi$, the $\ket{02}$ component of the two-phonon state ideally acquires a phase factor of $2\phi$ (in the linear limit). When the detuning between the two-phonon state and the scattering qubit is large compared to the phonon bandwidth, we can approximately achieve this linear phase shift, even though in general two-phonon
scattering is a nonlinear process \cite{Lund2023}. Following the phase shift induced by scattering from Q$_2$, interference at the BS produces the final output phonon state $\ket{\psi}_f=\sin{\phi}(\ket{02}+\ket{20})/\sqrt{2} + \cos{\phi} \ket{11}$.  We perform timed capture of the output of the BS with both qubits and measure the individual qubit excitation probabilities $P_\text{Q1,Q2}$ as well as the joint excitation $P_{ee}$ as a function of Q$_2$'s detuning. The time-domain process is shown in Fig.~\ref{fig3} \textbf{b} and \textbf{c} for two values of Q$_2$ detuning, and the dependence of the final qubit probabilities on Q$_2$ detuning is shown in panel \textbf{d}.  

The joint excitation $P_{ee}$ in panel \textbf{d} shows an interference pattern with a visibility of $\mathcal{V}_\text{ee}=0.968\pm0.004$. The two maxima of $P_{ee}$ correspond to a $2\phi=2\pi$ phase shift when $\ket{\psi}_f = \ket{11}$, while the minimum corresponds to a $2\phi=\pi$ phase shift when $\ket{\psi}_f = \ket{\psi}_\text{HOM}$ (see also Extended Data Fig.~\ref{figS4}). These results demonstrate the ability to jointly control the phase of one- and two-phonon states.

\begin{figure}[ht]
\begin{center}
	\includegraphics[width=0.8\textwidth]{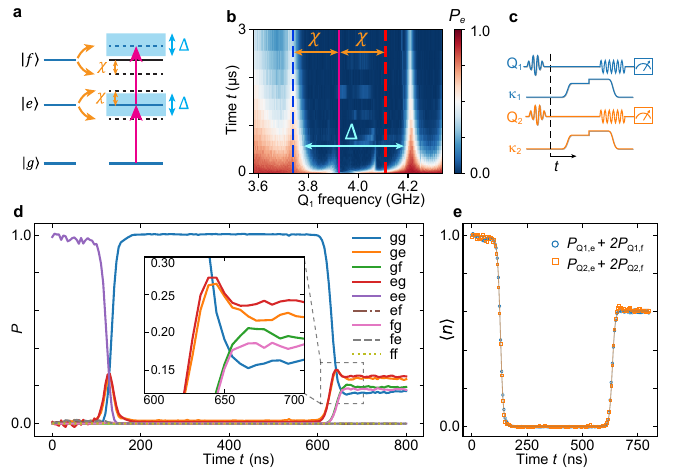}
	\caption{
    \label{fig4}
    {\bf Two-phonon coherent catch.}
\textbf{a,} Two-phonon catch concept. The qutrit $\ket{e}$ and $\ket{f}$ states are modulated by a microwave drive to create sidebands at $\pm \chi$, the qubit anharmonicity, yielding two transitions with equal energies (red arrows). $\Delta$ represents the acoustic transducer emission bandwidth. \textbf{b,} Q$_1$ phonon emission spectrum measured by Q$_1$ population decay as a function of frequency across the UDT bandwidth. The purple vertical line indicates the phonon center frequency at 3.925 GHz, while the red and blue dashed lines represent the sideband frequencies, located $\pm \chi = \pm 185$ MHz from the center frequency. \textbf{c,} Pulse sequence. Following dual phonon releases and interference at the BS, the two-phonon catch modulation is turned on while the coupling strength is adjusted during the catch process. \textbf{d,} Time domain release and coherent capture of the HOM output. Inset provides magnified view of the non-zero traces ($\ket{eg}$, $\ket{ge}$, $\ket{gf}$, $\ket{fg}$, $\ket{gg}$) from 600 to 700 ns. \textbf{e,} Expectation value of total excitation number for each qutrit as a function of time.}
\end{center}
\end{figure}

A challenge in manipulating and measuring two-phonon states is due to the intrinsic anharmonicity of the qubit, which makes it difficult to capture more than one phonon at a time (for example to capture a two-phonon HOM state). To overcome this, we propose and implement a sideband modulation protocol to enable a two-phonon catch, as illustrated in Fig.~\ref{fig4}{\bf a}. We set the qubit $g \leftrightarrow e$ transition frequency to match the incoming phonon frequency, enabling single-phonon capture. In addition, we modulate the capturing qubit with an external microwave tone at the qubit anharmonicity frequency $\chi=|\omega_{ef}-\omega_{ge}|$. This modulation generates sidebands at $\pm \chi$ for both the $g \leftrightarrow e$ and $e \leftrightarrow f$ transitions, yielding a non-zero matrix element for absorbing a two-phonon state, although also coupling other, unwanted transitions. The frequency-dependent coupling of the UDTs in our device allows us to place the $\ket{e} \leftrightarrow \ket{f}$ direct transition at a local maximum $T_1 = 1.5~\mu$s (blue dashed line in Fig.~\ref{fig4}{\bf b}), minimizing the unwanted $\ket{f}$ to $\ket{e}$ decay during the two-phonon capture process. 

To test this method, we first create a HOM state $\ket{\psi}_\text{HOM}$ by simultaneously interfering two phonons at the BS.  We then implement a two-phonon catch of the BS output by modulating both qubits Q$_{1,2}$ at their respective anharmonicity frequencies, $\chi_{1 (2)}=\omega_{ef}-\omega_{ge} = 185$ MHz (189 MHz). We experimentally optimize the modulation strength and coupler strength. We then perform joint qutrit state readout, yielding the time domain results shown in Fig.~\ref{fig4}\textbf{d}, showing non-zero probabilities for successful two-phonon capture, with final qubit probabilities $P(\ket{gf}) = 0.19$ and $P(\ket{fg}) = 0.18$. These accompany non-zero single-phonon captures, represented by the probabilities $P(\ket{ge}) = 0.24$ and $P(\ket{eg}) = 0.23$. The joint excitation probabilities for the $\ket{ee}$, $\ket{ef}$, $\ket{fe}$ and $\ket{ff}$ states are zero as expected.

In Fig.~\ref{fig4}{\bf e}, we show the expectation value $\langle n \rangle$ of the total excitation number for each qutrit, defined as $\langle n\rangle = P({\ket{e}}) + 2P({\ket{f}})$, also as a function of time. The final value of $\langle{n}\rangle_1=\langle{n}\rangle_2=0.61$ is approximately twice the $|e\rangle$ state probability of 0.32 observed for each qubit in the single-phonon experiments (Fig.~\ref{fig1}{\bf c}), suggesting minimal additional loss in the total phonon population for the two-phonon experiment compared to the single-phonon case.


In conclusion, we demonstrate one- and two-phonon phase gates, as well as phonon number-resolving detection, providing important tools for the advancement of phonon-based quantum computing. Our platform demonstrates high-visibility phonon interferometry, deterministic single-phonon sources with excellent phonon indistinguishability, and number-resolving phonon detection.  The phonon phase gates demonstrated here can likely be extended by using the qubit $e \leftrightarrow f$ transition in the scattering process \cite{Wang2024,Besse2020,Ferreira2024}. This will extend our quantum control of itinerant phonon qubits, and could find applications for e.g. quantum random access memories \cite{Wang2024} and quantum networks \cite{Kuzyk2018,Lemonde2018,SafaviNaeini2019,Neuman2021,Chen2023}.

\clearpage

\bibliographystyle{naturemag}
\bibliography{bibliography}

\clearpage


\subsection*{Methods}
\textbf{Single phonon scattering process.}
We provide an analytical model for the linear scattering process for a single itinerant phonon state from a transmon qubit. We assume the phonon state has frequency profile $u(\omega)=\int_{-\infty}^{\infty} u(t) \exp{-i\omega t} \text{d}t$, as implemented by $\ket{u}=\int_{-\infty}^{\infty} u(\omega) \opad_\text{in}(\omega) \ket{0} \text{d}\omega$. Upon scattering from the transmon qubit, which is coupled to the acoustic channel with maximum decay rate $\kappa_{\text{max}}$ and relative detuning $\Delta$ of the qubit $eg$ transition to the phonon center frequency, the  phonon state is reflected as $\ket{v}=\int_{-\infty}^{\infty} v(\omega) \opad_\text{out}(\omega) \ket{0} \text{d}\omega$, with frequency profile
\begin{equation}
    v(\omega) = u(\omega) \frac{i(\omega-\Delta) + \kappa_{\text{max}}/2}{i(\omega-\Delta) - \kappa_{\text{max}}/2} ,
\end{equation}
which can be derived by solving the input-output relations in the Heisenberg picture:
\begin{align}
    -i \omega \hat{a}(\omega) &= -i \Delta \hat{a}(\omega) - \frac{\kappa_\text{max}}{2} \hat{a}(\omega) - \sqrt{\kappa_\text{max}}\opa_\text{in}(\omega),\\
    \opa_\text{out}(\omega) &= \opa_\text{in}(\omega) + \sqrt{\kappa_\text{max}} \hat{a}(\omega).
\end{align}
Here $\opa_\text{in}(\omega)$ and $\opa_\text{out}(\omega)$ are the annihilation operators for the input and output phonon states.
The phase shift of output phonon state $\ket{v}$ and wave packet distortion are determined by the overlap
\begin{equation}
    \bra{u}\ket{v} = \int_{-\infty}^{\infty} |u(\omega)|^2 \frac{i(\omega-\Delta) + \kappa_{\text{max}}/2}{i(\omega-\Delta) - \kappa_{\text{max}}/2} \text{d}\omega .
\end{equation}
The phase shift is given by $\arg(\bra{u}\ket{v})$. Distortion of the phonon waveform is minimized ($\abs{\bra{u}\ket{v}}\rightarrow 1$) when either $\kappa$ or $\Delta$ is significantly larger than the bandwidth of the input phonon waveform. Given that in these experiments, the phonon temporal waveform $u(t) \propto \sech(t/\sigma_{1,2})$, we plot the expected scattering phase $\arg{\bra{u}\ket{v}} $ and pulse distortion $1-\abs{\bra{u}\ket{v}}$ as a function of detuning frequency in Extended Data Fig. \ref{figS7}. The expected scattering phases at $\arg{\bra{u}\ket{v}} = \pi/2$ and $3\pi/2$ align with the detuning frequency for the two experimental single-phonon excitation minima, corresponding to a net $\pi$ phase difference between these points. We estimate the pulse distortion to be at the level of $3 \times 10^{-2}$ to $10^{-3}$, depending on the detuning.

\textbf{Theory for two-phonon absorption using qubit modulation.}
The transmon Hamiltonian with modulation frequency $\Omega$ and modulation amplitude $\delta$ is given by
\begin{equation}
    \oph_0(t) = (\omega_0 + \delta \cos (\Omega t)) \opad \opa - \frac{\chi}{2} \opad \opad \opa \opa .
\end{equation}
We can transform to the interaction frame via the unitary
\begin{equation}
    \begin{split}
        \opu(t) =& \exp \left( -i \int_0^t \oph_0(t') dt' \right) \\
        =& \exp \left( i \frac{\chi t}{2} \opad \opad \opa \opa \right) \exp \left( -i \opad \opa \left( \omega_0 t + \frac{\delta}{\Omega} \sin \Omega t \right) \right) .
    \end{split}
\end{equation}
The qubit operator $\opa$ in the interaction frame becomes the time-dependent operator $\opa(t)$,
\begin{equation}
    \begin{split}
        \opa(t) =& \opud (t) \opa \opu(t) \\
        =& \opud (t) (\ket{g}\bra{e} + \sqrt{2} \ket{e}\bra{f} + ...) \opu(t) \\
        =& \exp \left( -i \left( \omega_0 t + \frac{\delta}{\Omega} \sin \Omega t \right) \right) \left( \ket{g}\bra{e} + \sqrt{2} e^{i\chi t} \ket{e}\bra{f} + ... \right) \\
        =& e^{-i \omega_0 t} \left( \ket{g}\bra{e} + \sqrt{2} e^{i\chi t} \ket{e}\bra{f} + ... \right) \sum_{-\infty}^{\infty} J_n \left( \frac{\delta}{\Omega} \right) e^{-i n \Omega t} .
    \end{split}
\end{equation}
If we modulate the qubit at the anharmonicity $\Omega = \chi$ and drop terms not at $\omega_0$, we are left with the constant terms in the rotating frame
\begin{equation}
    \opa (t) = J_0 \left( \frac{\delta}{\Omega} \right) \ket{g}\bra{e} + J_1 \left( \frac{\delta}{\Omega} \right) \sqrt{2} \ket{e}\bra{f} + \ldots
\end{equation}
To make the system behave like a simple harmonic oscillator, we impose the condition
\begin{equation}
    J_0 \left( \frac{\delta}{\Omega} \right) = J_1 \left( \frac{\delta}{\Omega} \right),
\end{equation}
which can be achieved by setting the modulation amplitude $\delta \approx 1.43 \chi$.

\textbf{Qubit measurement correction.}
A qubit measurement correction \cite{Steffen2006, Bialczak2010} is applied to all the qubit population data except that in Fig. \ref{fig1}\textbf{d} and Extended Data Figs. \ref{figS1}, \ref{figS2}, \ref{figS4} and \ref{figS7}\textbf{a}. All reported visibilities are calculated from uncorrected data.


To calculate the two-qubit measurement corrections, we measure both qubits simultaneously using a multiplexed readout pulse. Prior to each experiment, we measure the two-qubit readout visibility matrix, by preparing the two qubits in the fiducial states $\qty\big{\ket{gg}, \ket{ge}, \ket{eg}, \ket{ee}}$, followed by two-qubit readout. The visibility matrix $V$ is defined as the linear transformation between the measured probability vector and the expected probability vector for the different fiducial states, $P_{meas} = V P_{exp}$. A typical visibility matrix is:
\begin{equation*}
    V=\begin{pmatrix}
        F_{gg,gg} & F_{gg,ge} & F_{gg,eg} & F_{gg,ee}\\
        F_{ge,gg} & F_{ge,ge} & F_{ge,eg} & F_{ge,ee}\\
        F_{eg,gg} & F_{eg,ge} & F_{eg,eg} & F_{eg,ee}\\
        F_{ee,gg} & F_{ee,ge} & F_{ee,eg} & F_{ee,ee}
    \end{pmatrix} = 
    \begin{pmatrix}
        0.988(1)& 0.006(1)& 0.006(1)& 0.00002(6)    \\
        0.050(3)& 0.944(4)& 0.0002(2)& 0.006(1) \\
        0.082(3)& 0.0005(2)& 0.912(3)& 0.005(1) \\
        0.005(1)& 0.090(3)& 0.045(3)& 0.861(5) \\
    \end{pmatrix}
\end{equation*},
where $F_{a,b}$ represents the fidelity of preparing the two-qubit state $\ket{a}$ but measuring the two-qubit state $\ket{b}$. By inverting the visibility matrix, we obtain the measurement-corrected two-qubit probability vector $P_{corr} = V^{-1} P_{meas}$.

For the two-qutrit measurement correction, we separately measure the visibility matrix for each qutrit, and define the two-qutrit visibility matrix as the the tensor product of these individual matrixes. Hence the two-qutrit visibility matrix used for the data in Fig.~\ref{fig4}{\bf d} is given by
\begin{align*}
    V&=\begin{pmatrix}
        F_{1,g,g} & F_{1,g,e} & F_{1,g,f}\\
        F_{1,e,g} & F_{1,e,e} & F_{1,e,f}\\
        F_{1,f,g} & F_{1,f,e} & F_{1,f,f}
    \end{pmatrix} \otimes
    \begin{pmatrix}
        F_{2,g,g} & F_{2,g,e} & F_{2,g,f}\\
        F_{2,e,g} & F_{2,e,e} & F_{2,e,f}\\
        F_{2,f,g} & F_{2,f,e} & F_{2,f,f}
    \end{pmatrix} \\
    &=\begin{pmatrix}
        0.983(1)& 0.017(1)& 0.0003(2)   \\
        0.104(9)& 0.884(9)& 0.011(1)\\
        0.028(3)& 0.108(7)& 0.864(7)
    \end{pmatrix}\otimes
    \begin{pmatrix}
        0.988(1)& 0.010(1)& 0.0016(6)\\
        0.10(3)& 0.90(3)& 0.003(2)\\
        0.020(3)& 0.079(4)& 0.901(5) \\
    \end{pmatrix},
\end{align*}
where $\otimes$ represents the tensor product. Parenthetical numbers represent uncertainties.
\clearpage

\section*{Supplementary Information}
\setcounter{figure}{0}
\renewcommand{\figurename}{\textbf{Extended Data Figure}}

\begin{figure}[ht]%
\centering
\includegraphics[width=0.9\textwidth]{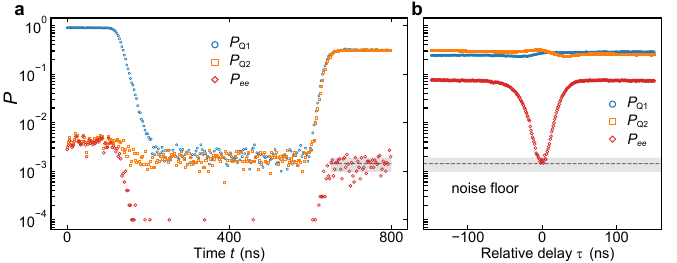}
\caption{\textbf{Noise floor of HOM experiment.} \textbf{a,} Data shown in Fig.~\ref{fig1}b in log scale, here without visibility correction (see Methods). For the single-phonon splitting experiment, we expect at most one excitation during the entire process. The ideal $P_{ee}$ is zero for the duration of the experiments, and in particular, it should be zero after 650 ns, when each qubit has completed its capture of the split single-phonon wavepacket. We attribute the non-zero $P_{ee}$ population after that time to residual qubit thermal excitations and readout errors. We display the mean $P_{ee}$ population $P_{ee\text{,floor}}=0.0014 \pm 0.0005$ from 650 to 800 ns (dashed gray line, with uncertainty represented by the shaded area). We note the suppression in $P_{ee}$ between phonon release and capture,  for times between 200 and 600 ns with $P_{ee} \sim 10^{-4}$, is likely due to cooling of the qubits via the acoustic channel, showing a significant portion of the residual $P_{ee}$ is attributable to non-thermal excitations. \textbf{b,} Data shown in Fig.~\ref{fig1}d together with $P_{Q1}$ and $P_{Q2}$ in log scale, again here without visibility correction. The two-phonon interference experiments use a pulse sequence similar to that for the single-phonon experiments, and the final populations for $P_{Q1}$ and $P_{Q2}$ are close to those for the single-phonon experiments \cite{Qiao2023}. We observe $P_{ee}$ is suppressed near $\tau=0$, and at $\tau = 0$ it is consistent with the baseline noise floor $P_{ee,floor}$. We thus believe  $P_{ee\text{,floor}}$ is a reasonable estimate for the noise floor in this experiment. 
}\label{figS1}
\end{figure}

\begin{figure}[ht]%
\centering
\includegraphics[width=0.9\textwidth]{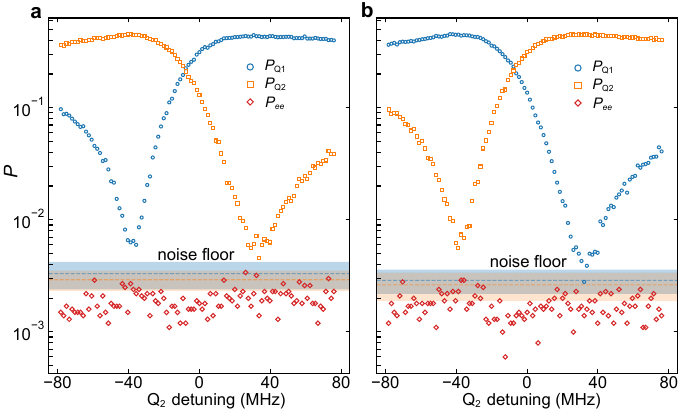}
\caption{\textbf{Uncorrected single phonon phase gate data.} \textbf{a, b,} Data from Fig.~\ref{fig2}c and d plotted here in log scale, without a visibility correction. The average visibility for the Mach-Zehnder interferometry (not to be confused with the measurement visibility matrix) is $\mathcal{V}_\text{MZ,raw}\equiv (P_\text{max}-P_\text{min})/(P_\text{max}+P_\text{min})=0.979 \pm 0.006$. We also measured the excited state population for each qubit after releasing a single phonon and being cooled by the acoustic channel, shown by the corresponding colored dashed lines, with the color-shaded regions representing one standard deviation in uncertainty. The noise floor limits the maximum measurable visibility to $(P_\text{max}-P_\text{floor})/(P_\text{max} + P_\text{floor}) =  0.986 \pm 0.001$.
}\label{figS2}
\end{figure}

\begin{figure}[ht]%
\centering
\includegraphics[width=0.9\textwidth]{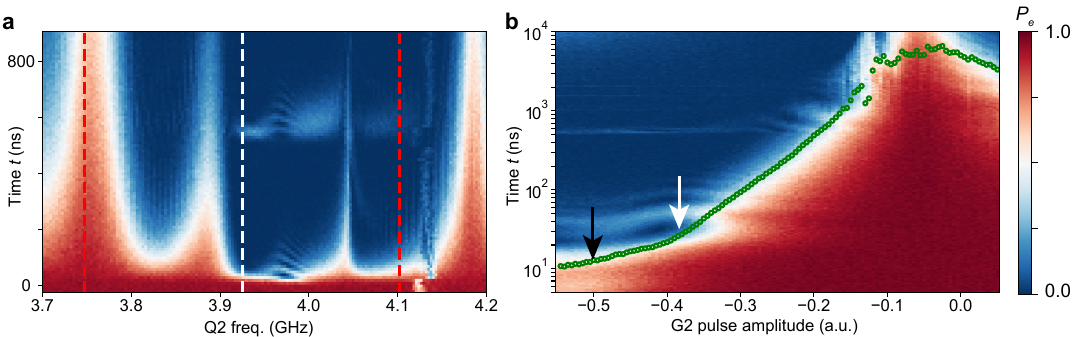}
\caption{\textbf{Characterization of the scattering qubit Q$_2$ acoustic emission spectrum.} \textbf{a,} Qubit $\text{Q}_2$'s emission spectrum, measured by exciting the qubit and monitoring its subsequent excited state population $P_e$ (color scale) as a function of time (vertical scale) as the qubit frequency is tuned across the UDT bandwidth (horizontal scale). The beamsplitter reflection yields features at $\sim 550$ ns, visible within the transducer unidirectional band from 3.91 GHz to 4.05 GHz. White dashed line indicates the qubit operating frequency of 3.925 GHz. Red dashed lines show the qubit $\pm \chi = \pm 189$ MHz sideband frequencies when modulating the qubit at that frequency. \textbf{b,} Tunable qubit-UDT excited state population (color scale) as a function of time (vertical scale) and coupler $\text{G}_2$ control pulse amplitude (arbitrary units; horizontal scale), for the initially excited-state qubit set to the operating frequency of 3.925 GHz. Green dots are fit $T_1$ times as a function of G$_2$ pulse amplitude. White arrow indicates coupling strength used in panel {\bf a}, and used for single-phonon release from Q$_2$. Black arrow indicates maximum coupling strength used in the phonon scattering experiments.
}\label{figS3}
\end{figure}

\begin{figure}[ht]%
\centering
\includegraphics[width=0.9\textwidth]{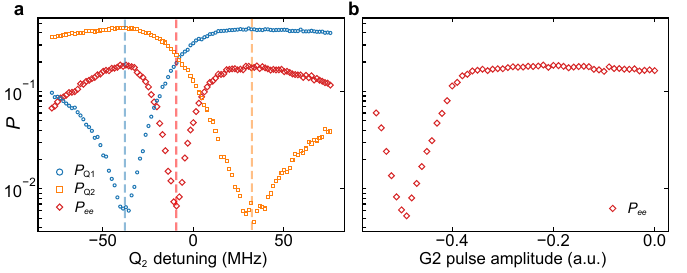}
\caption{\textbf{Two-phonon phase gate measurements.} \textbf{a,} Two-qubit joint excitation probability $P_{ee}$ as a function of Q$_2$ scattering detuning, as shown in Fig.~\ref{fig3}d, and here including the single-phonon phase gate data shown in Fig.~\ref{fig2}c, in log scale without visibility matrix correction. The two maxima in $P_{ee,\text{max}}$ align in Q$_2$ frequency with the single qubit excitation minima $P_{Q1,\text{min}}$ and $P_{Q2,\text{min}}$ (blue and orange dashed line), indicating the two-phonon phase shift is approximately twice that of the single-phonon phase shift. \textbf{b,} For fixed Q$_2$ detuning at -9.2 MHz (red dashed line in panel \textbf{a}), the two-qubit joint excitation $P_{ee}$ is shown as a function of the coupler $\text{G}_2$ control pulse amplitude. The minimum $P_{ee,\text{min}}$ is reached for a pulse amplitude of -0.5 (a.u.), corresponding to maximum coupling. The coupler is turned off when the pulse amplitude is near zero, with no phase shift applied, where $P_{ee}$ approaches its maximum. The average visibility $\mathcal{V}_\text{ee,raw} \equiv (P_{ee,\text{max}}P_{ee,\text{min}})/P_{ee,\text{max}} = 0.968 \pm 0.004$. }\label{figS4}
\end{figure}

\begin{figure}[ht]%
\centering
\includegraphics[width=0.9\textwidth]{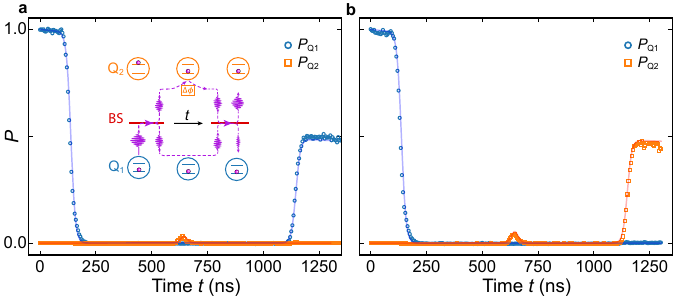}
\caption{\textbf{Time domain single-phonon phase gate.} \textbf{a, b,} As a supplement to Fig.~\ref{fig2}a and b, when a single phonon is released by Q$_1$ and scattered by Q$_2$, the final output of the beamsplitter is routed towards $\text{Q}_1$ or $\text{Q}_2$, depending on the phase introduced by scattering from Q$_2$. Inset in panel \textbf{a} shows the schematic of the experiment.
}\label{figS5}
\end{figure}

\clearpage
\begin{figure}[ht]%
\centering
\includegraphics[width=0.9\textwidth]{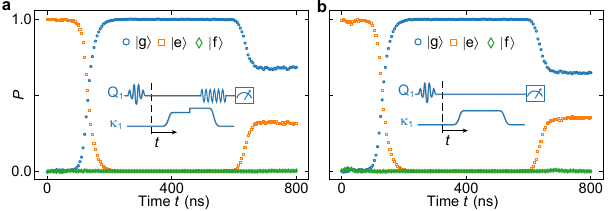}
\caption{\textbf{Control experiment with and without qubit modulation.} Each panel shows the measured qutrit population with time for two different catch protocols. Inset pulse sequences illustrate the differences between the two experiments. \textbf{a,} a single phonon is released from Q$_{1}$ and partially reflected by the BS. During the subsequent catch process, we use the qubit modulation pulse sequence shown in Fig.~\ref{fig4}\textbf{c}.  \textbf{b,} Catch process using the time-reversed control for the release process. Both experiments yield similar catch efficiencies (0.32 vs. 0.34). We observe no spurious phonon excitation to the transmon $\ket{f}$ state, with or without qubit modulation.  
}\label{figS6}
\end{figure}

\begin{figure}[ht]%
\centering
\includegraphics[width=0.7\textwidth]{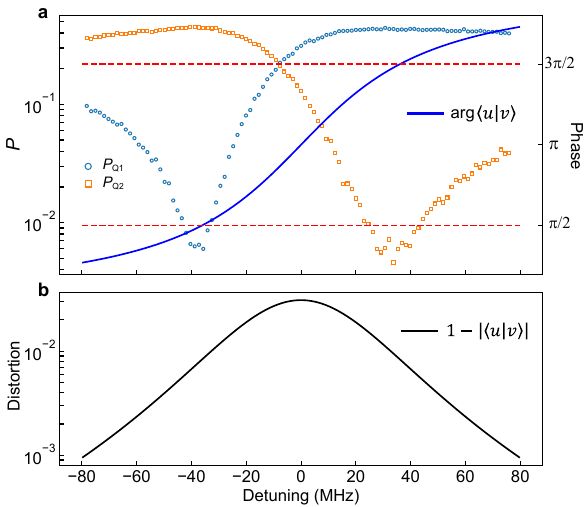}
\caption{\textbf{Theoretical phonon scattering process.} \textbf{a,} Theoretical scattering phase $\arg{\bra{u}\ket{v}}$ (blue line; right vertical axis) as a function of detuning frequency, together with single-phonon phase gate data shown in Fig.~\ref{fig2}c (orange and aqua points; left vertical axis). The theoretical scattering phases at $\arg{\bra{u}\ket{v}}= \pi/2$ and $3\pi/2$ align in detuning frequency with the two experimental single-phonon excitation minima, corresponding to a net $\pi$ phase difference. \textbf{b,} Theoretical pulse distortion $1-\abs{\bra{u}\ket{v}}$ as a function of detuning frequency. The pulse distortion decreases from $3\times10^{-2}$ at zero detuning to $6\times10^{-3}$ at the detuning where $\arg{\bra{u}\ket{v}}= \pi/2$ and $3\pi/2$.} \label{figS7}
\end{figure}

\clearpage

\begin{table}[H]
\centering
\begin{tabular}{@{}llll@{}}
\toprule
Qubit parameters & Qubit 1  & Qubit 2 \\
\midrule
Idle frequency (GHz)    & 4.206   & 4.072\\
Anharmonicity (MHz)  &-186 & -190\\
Intrinsic lifetime $T_1$ ($\mu s$)   & 33.5   & 28.2\\
Readout resonator frequency (GHz)  & 5.139   & 5.073 \\
$\ket{e}$ state visibility  & 0.952   & 0.913 \\
$\ket{g}$ state visibility  & 0.998   & 0.998 \\
$\ket{f}$ state visibility  & 0.864   & 0.901 \\
\midrule
SAW parameters & Beamsplitter  & Mirror & IDT  \\
\midrule
Number of cells (design)  &14&400&28 \\
Aperture ($\mu m$) (design)  &150&150&150 \\
Pitch ($\mu m$) (design) &0.5 & 0.5 & 0.975\\
Metallization ratio (SEM) &0.72&0.72& 0.50 \\
\botrule
\end{tabular}
\captionsetup{labelformat=empty}
\caption{Extended Data Table.~1: \textbf{Summary of device parameters.} Qubit lifetimes, anharmonicities, and state visibilities are measured at the qubit idle frequencies. One cell in each SAW element is the smallest repeating unit; the SAW aperture is the width measured perpendicular to the SAW propagation direction; the pitch is the center-to-center finger distance, and the metallization ratio is the ratio of finger metal width to gap width. Metallization is 30 nm thick aluminium, patterned by electron beam lithography lift-off.}
\end{table}

\subsection*{Data availability}
The data display in figures and other findings of this study are available from the corresponding author upon reasonable request.

\subsection*{Acknowledgements}
This work was supported by the Defense Advanced Research Projects Agency (DARPA) under Agreement No. HR0011-24-9-0364, the Air Force Office of Scientific Research (AFOSR grant FA9550-20-1-0270), AFOSR MURI (FA9550-23-1-0338), by UChicago's MRSEC (NSF award DMR-2011854) and by the NSF QLCI for HQAN (NSF award 2016136). This material is based upon work supported by the U.S. Department of Energy, Office of Science, National Quantum Information Science Research Centers. This work was partially supported by UChicago's MRSEC (NSF award DMR-2011854) and by the NSF QLCI for HQAN (NSF award 2016136). We made use of the Pritzker Nanofabrication Facility, which receives support from SHyNE, a node of the National Science Foundation's National Nanotechnology Coordinated Infrastructure (NSF Grant No. NNCI ECCS-2025633). Correspondence and requests for materials should be addressed to A. N. Cleland (anc@uchicago.edu). 

\subsection*{Author contributions}
H.Q. designed and fabricated the devices, performed the measurements and analyzed the data. Z.W. developed the theoretical protocol for two-phonon catch. A.N.C. advised on all efforts. All authors contributed to the discussions and production of the manuscript.

\subsection*{Competing interests}
The authors declare no competing interests.

\end{document}